\documentclass[a4paper,aps,prb,twocolumn,amsmath,amssymb,groupedaddress]{revtex4}

\usepackage{dcolumn}
\usepackage{graphicx,subfigure}
\usepackage{bm}
\usepackage{verbatim}
\usepackage{amsmath}
\usepackage{amssymb}
\usepackage[T1]{fontenc}
\usepackage{ae,aecompl}
\usepackage{appendix}
\usepackage{float}
\usepackage{color}
\usepackage[export]{adjustbox}
\usepackage{array}

\newcommand{\rv}{{\bf r}}

\newcommand{\kv}{{\bf k}}

\newcommand{\be}{\begin{equation}}
\newcommand{\ee}{\end{equation}}
\newcommand{\bea}{\begin{eqnarray}}
\newcommand{\eea}{\end{eqnarray}}
\newcommand{\bse}{\begin{subequations}}
\newcommand{\ese}{\end{subequations}}

\begin{document}
\title{High-Performance Photonic Crystal Biosensors using Sub-wavelength Nanopillars}
\author{Zhengzheng Zhai and Sajeev John}
\affiliation{
Department of Physics, University of Toronto, Toronto, Ontario, Canada M5S 1A7}
\date{09/02/2026}
\email{zhengzheng.zhai@utoronto.ca}

\begin{abstract}
Photonic-crystal (PC) slabs composed of subwavelength silicon pillars erected on a thin silicon backing layer, with heights of only one or half lattice constant, are simulated for optical biosensing by numerical solution of Maxwell’s equations. Localized optical cavity modes that form waveguides in direction of biofluid flow, with quality factors surpassing $10^4$, are embedded within 2D photonic band gap of the 3D structure. Remarkable sensitivity to analyte binding is achieved using fragmented thin silicon pillars in regions of maximum localized optical-field intensity. Exceptional quality factors are realized through destructive wave interference of vertical radiation loss. This provides a rare combination of high sensitivity and low limit of detection in a device with minimal time requirement for analyte binding. The sensitivity to analyte binding is a frequency shift of nearly $0.004\; [2\pi c/a]$ per analyte-thickness increment $0.1a$. The sensitivity to background biofluid change is about $4800$ nm per refractive index unit (RIU) for a specific lattice constant 
$a = 5\;\mu$m using light of wavelength $16.5\; \mu$m.
\end{abstract}
\pacs{}

\maketitle
Optical biosensors\cite{Damborsky16,Lee07,Chow04,Skivesen07,Inan17} based on localized electromagnetic resonances are a powerful platform for
rapid, label-free biochemical detection. Nevertheless, high-performance biosensors with both high sensitivity and low limit-of-detection have remained elusive. It is challenging to simultaneously achieve a high optical ﬁeld concentration in the analyte region and very high quality ($Q$) factor of the resonance. In conventional biosensors, these two requirements are often in conflict \cite{Mohamed10,SPR}.  Photonic crystal (PC) biosensors \cite{Abdullah15,Abdullah19,Dragan21,Zhai2026} offer a fundamental path to achieving both. They provide spatially localized defect and waveguide modes within the ``blank slate" of complete 2D photonic band gap (PBG). Such engineered optical resonances can be well separated in
frequency from spurious resonances and provide a ``broad net" for the capture of analytes for rapid detection and diagnosis.

Sensitivity is defined as
the rate of change of a resonant mode frequency with either increment of attached analyte thickness for fixed analyte refractive
index or change in overall background biofluid refractive-index due to unattached analyte. The limit of detection is defined as the minimum detectable variation in analyte coating thickness or refractive index, defined by the spectral linewidth of the resonance. The frequency shift of the resonance must be comparable to or larger than the resonance linewidth for unambiguous detection. High-performance biosensors require simultaneously large sensitivity and high-\(Q\) resonances. In this paper, we show that this is achievable in a real-world 3D silicon-based photonic crystal biosensor through a combination of structural fragmentation and engineered asymmetry (period doubling) along the direction of defect waveguides within the PBG.

Typical high-sensitivity defect resonances exhibit a intrinsically low $Q$-factor even when the resonant frequency lies deep inside the in-plane photonic bandgap \cite{Zhai2026,ZhaiPB26}. Alternatively, high-$Q$ resonances exhibit weak sensitivity \cite{Dragan21,ZhaiPB26}. The cause of the low-$Q$ behavior in 3D is strong out-of-plane radiation loss. In photonic-crystal slabs, there are two decay mechanisms, and the total quality factor can be decomposed as \cite{PC-book08}$Q^{-1}=Q^{-1}_{\parallel}+Q^{-1}_{\perp}$, where \(Q_{\parallel}\) describes in-plane confinement by the 2D PBG (limited by waveguide decay into the cladding glass on either side of the biofluid flow channel), while \(Q_{\perp}\) is limited by vertical radiation leakage. We defined a coordinate system in which light transmission across the flow channel is the $x$-direction, biofluid flows along the $y$-direction and radiation into free space occurs in the $z$-direction. Silicon pillars, constituting the PC with lattice constant, $a$, are arranged within the flow channel and defect waveguides extend along the $y$-direction, but confine light in the $x$-direction (see Fig. \ref{fig: H=1a}). Increasing the chip width in the light-propagation $x$-direction primarily suppresses in-plane leakage and increases \(Q_{\parallel}\), whereas \(Q_{\perp}\) remains largely unchanged. Consequently, the total quality factor typically saturates to the intrinsic radiation limit \(Q_{\perp}\sim600\text{--}800\). \cite{ZhaiPB26}

The strong radiation loss originates from the fragmented defect geometries required for high sensitivity. 
For the simplest localized light mode within the 2D PBG, the optical Fourier components are peaked about $\kv_\parallel=(\pm \pi/a,0)$ in the in-plane wavevector $\kv_{\parallel}=(k_x, k_y)$ space. The fragmented thin strip geometry broadens the spectrum in $\kv$-space about their peaks and generates substantial Fourier components inside the light cone. This enhances coupling to free-space radiation channels\cite{Multipole19}. Consequently, even though the resonances are well confined in-plane by the 2D PBG, strong vertical leakage limits the achievable quality factor.

In this work, we introduce transverse period doubling along the waveguide line defect with fragmented thin silicon, allowing incident light with in-plane wavevector $(k_x,0)$ to couple to more efficacious optical waveguide modes with minimal vertical radiation loss. This enables simultaneous strong analyte-field interaction and very low radiation loss. These otherwise ``invisible" waveguide modes bear some similarities to the much touted ``quasi-bound states in the continuum" \cite{BIC16,BIC23,BIC23-2}. 

Our biosensor platform consists of a silicon photonic-crystal slab (black) embedded between glass cladding layers (blue) that constitutes a flow channel for biofluid, as shown in Fig. \ref{fig: H=1a} (a). A thin silicon backing layer is placed in between the PC structure and a glass substrate with refractive index $n=1.5$  ($\varepsilon=2.25$) to enhance the vertical light confinement. The photonic-crystal slab is composed of periodically arranged short silicon pillars with high refractive index $n=3.4$  ($\varepsilon=11.56$) immersed in a biofluid environment of $n_\mathrm{fluid}=1.35$  ($\varepsilon=1.8225$). This strong refractive-index contrast produces a 2D photonic band gap (PBG) in which light transmission is prohibited except by evanescent wave tunnelling through engineered defect resonances. We choose the optimal ratio of the width $w$ of the silicon block to the lattice constant $a$ to be $w/a=0.40$, in order to maximize the PBG. In the first illustration, pillar height is set to be $\mathrm{H}=1a$, while the surrounding glass wall of thickness $2a$ at both chip ends (flow channel sidewalls) is higher than the pillars by one lattice constant $a$ in order to guide the biofluid flow. The period-doubled line defect is placed in the middle of the PC slab as shown in Fig. \ref{fig: H=1a} (a). This specific design has only 2 unit cells of PC on either side of the defect region. Increasing the number of unit cells in the $x$-direction can greatly enhance the $Q$-factor as shown in Fig. \ref{fig: H=1a} (c). With period-doubling in the $y$-direction, the in-plane PC structure can be regarded as two sublattices A and B with unit cell of size $(1a \times 2a)$ in the $xy$ plane. The dimensions of each thin silicon pillar are $\left(w_s \times (h\pm dh) \times \mathrm{H}\right)=\left(0.22a \times (0.8\pm 0.1)a \times 1a\right)$, where $w_s=0.22a$ is the width in the $x$-direction, $(h\pm dh)=(0.8\pm 0.1)a$ are the different $y$-direction lengths. The spacing between two nearest thin pillars within the defect region is $w_o=0.2a$. This structure is periodically repeated in the $y$-direction. 

A normally incident Gaussian pulse is emitted by the light source (depicted as a red line segment) embedded in the left-side glass. While some light diffracts above and below the PC, the reminder evanescently tunnels in the $x$-direction through the PBG and the transmitted light is collected by the detectors (depicted as yellow line segments) embedded in the right-side glass.  The excitation source spans the full pillar height, with a cross-sectional area of 
$2a \times 1a$ in the yz-plane, ensuring uniform excitation across the silicon nanopillar. A fictitious, numerical detector spanning the same size as the light source is placed slightly after the light source (just $0.1$a away from the source), to evaluate the total source power before the light enters into the PC structure. To collect the transmitted light, three flux detectors (depicted as another three yellow lines) are positioned in the right-side glass region beyond the PC structure. Each detector has an area of $2a\times(a/3)$ in the yz-plane, corresponding to one third of the pillar height. The detectors are vertically stacked and labeled
$z(-1)$, $z(0)$, and $z(1)$, from bottom to top, with $z(0)$ denoting the central detector aligned with the mid-plane of the pillar. The total transmitted flux is obtained by summing the contributions from these three detectors, and the transmission is evaluated by dividing this sum by the total source power detected slightly after the light source.

The electric field ${\bf E}(\rv,t)$ emitted from the optical source is polarized in the vertical $z$-direction. In all the electric field profiles of the resonant modes, shown in Fig. \ref{fig: H=1,modes and binding}, we consider only the $\mathrm{E}_z$ component both in the $xy$ plane and the $xz$ plane. The biofluid flows in the $y$-direction. We apply periodic Bloch boundary conditions (PBCs) in the $y$-direction. Absorbing boundary conditions are imposed in the $x$- and $z$-directions using perfectly matched layers
(PMLs) to avoid spurious light reflection. The PMLs of thickness $3a$ are placed just behind both the glass walls at two chip ends, and also above the chip structure and below the glass substrate. We perform an FDTD simulation of total optical transmission through the chip using MEEP\cite{Meep}. The spatial resolution implemented for the FDTD calculations is 12 mesh points per lattice constant in each direction with subpixel smoothing of the dielectric function $\varepsilon(\rv)$\cite{Meep}.

The measured transmission generally consists of two contributions: (i) a broadband, non-resonant background component associated with direct or continuum propagation channels above and below the photonic crystal region, and (ii) a narrowband resonant component enabled by coupling through the localized defect mode. The interference between these two fields produces a characteristic Fano line shape\cite{Fan02,Fano14,Coupled-mode} for the resonance transmission peaks. These two contributions exhibit markedly distinct temporal dynamics: The broadband background decays rapidly after the excitation pulse has traversed the structure, whereas light coupled to the localized defect mode with frequency $\omega_0$  decays exponentially as $\mathrm{E}_{\text{res}}(t) \propto e^{-t/(2\tau)}\cos(\omega_0t)$, where the lifetime $\tau$ is related to the quality factor by $Q = \omega_0 \tau$. To isolate the signal passing through the PBG, we apply time-delay filtering. Using a delay time $t_{\text{delay}}$, we retain only the late-time portion of the transmitted field \cite{ZhaiPB26}. For Chip I, the delay time is chosen to be  $t_\mathrm{delay} =150\; [a/c]$ (equal to $2.5 \times 10^{-12}s$ for a specific unit-cell size $a=5\;\mu$m), and the resulting transmission spectrum is shown in Fig. \ref{fig: H=1a} (b). With continuum background eliminated, each previous Fano lineshape becomes a Lorentzian lineshape.

\begin{widetext}
	\begin{center}
		\begin{figure}[htbp]
			\subfigure[]{   \hspace{-0.1in}\raisebox{-0.1in}{\includegraphics[width=0.92\textwidth]{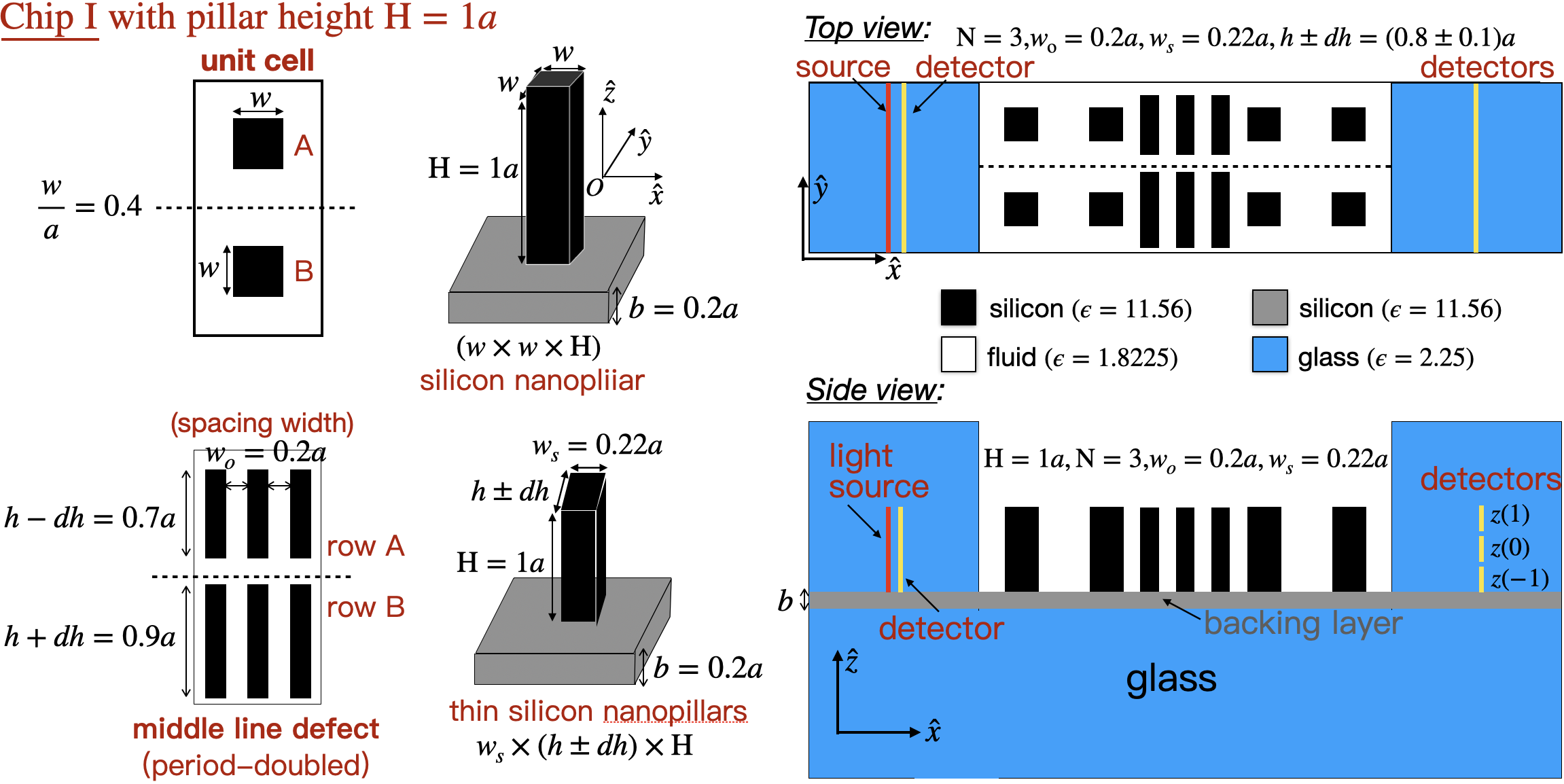}}}
			\hspace{0in}
			\subfigure[]{
				\hspace{-0.3in}\raisebox{-0.1in}{\includegraphics[width=0.48\textwidth]{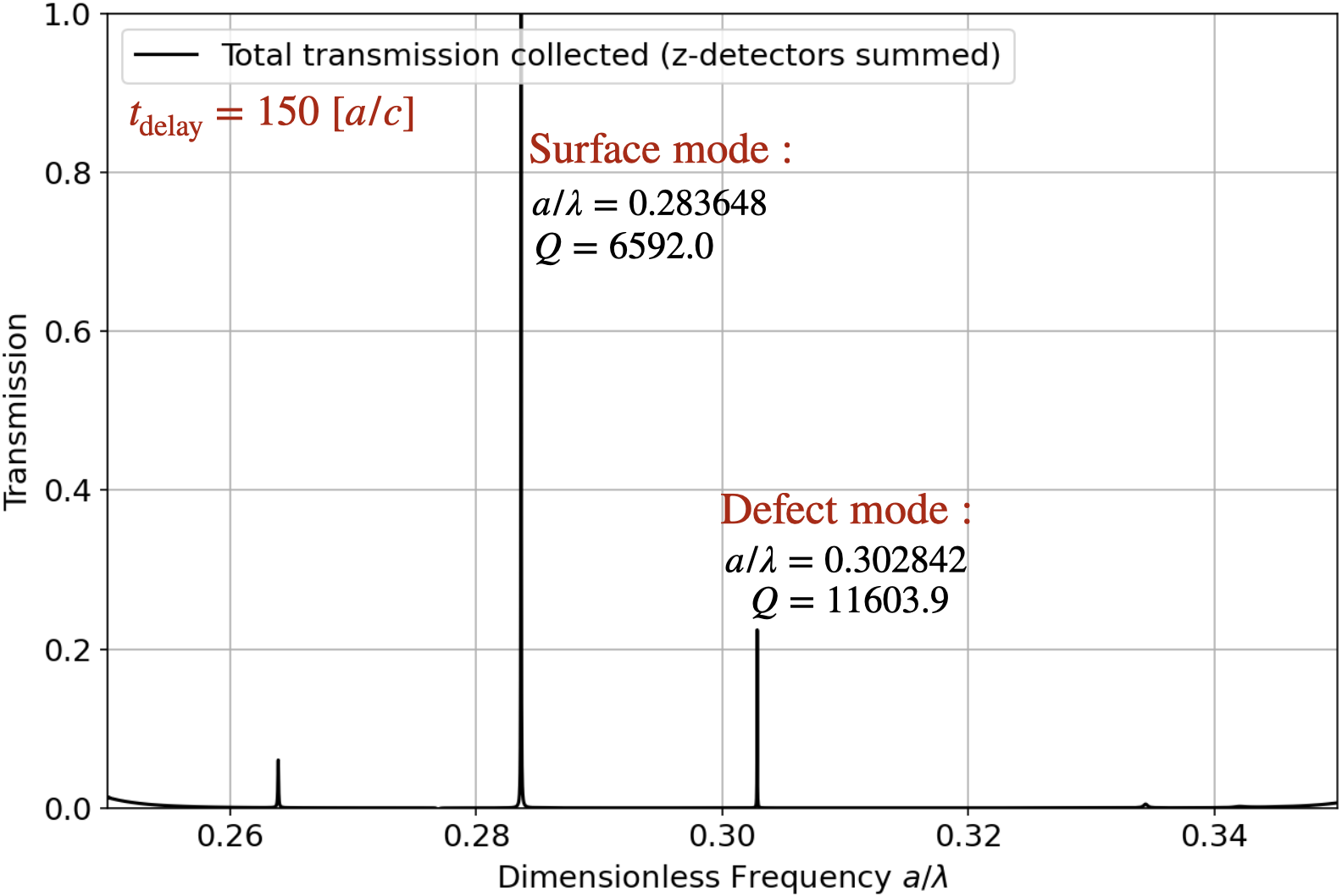}}}
			\hspace{0.15in}
			\subfigure[]{\raisebox{-0.1in}
				{\includegraphics[width=0.49\textwidth]{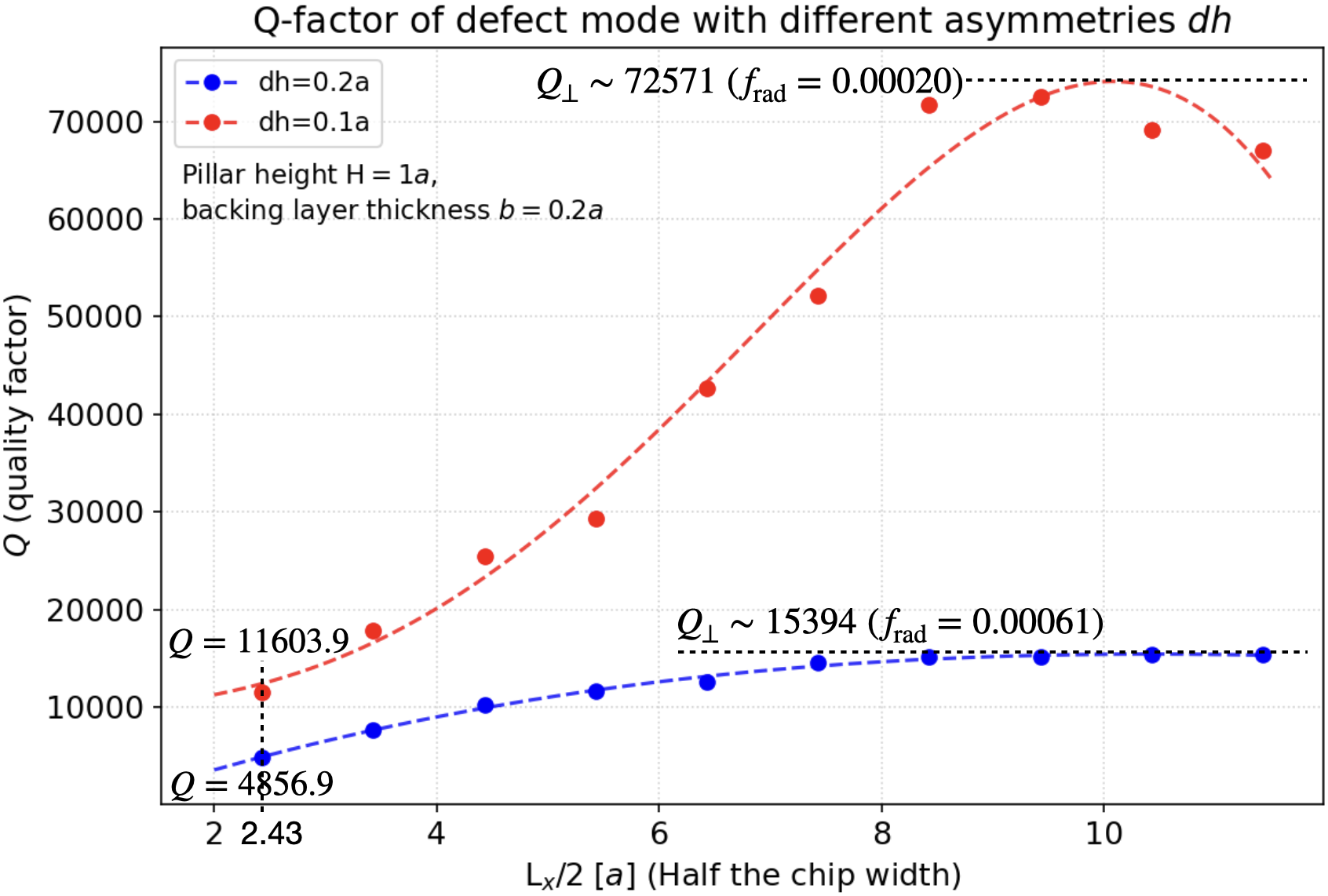}}}
			\caption{Chip I with pillar height $\mathrm{H}=1a$. (a) A period-doubled line defect in the middle of the PC slab placed upon a thin silicon backing layer of thickness $b=0.2a$. The in-plane cross-section of each thin silicon nanopillar in the defect region is $\left(w_s \times (h\pm dh)\right)=\left(0.22a \times (0.8\pm 0.1)a \right)$. The spacing between two nearest thin nanopillars within the defect region is $w_o=0.2a$. The spatial separation between the center of the defect region and the nearest regular silicon block is $0.93a$. The $y=0$ plane (dashed line in the top view) lies in the middle of the two sublattices A and B in the biofluid flow direction. (b) Transmission spectrum after time-delay filtering with a delay time $t_\mathrm{delay}=150\;[a/c]$. With a small asymmetry $dh=0.1a$, the defect mode has a very high quality factor $Q=11603.9$, even with a short chip width $\mathrm{L}_x=4.86a$, with only two silicon blocks in between the defect region and the chip end. (c) The $Q$-factor of the defect mode saturates to its intrinsic radiation limit $Q_\perp \sim 72571$ as the chip width $\mathrm{L}_x$ in the light propagation ($x$) direction is increased to a large value. Such intrinsic radiation limit is controlled by the vertical radiation loss alone and therefore related to the radiation fraction $f_\mathrm{rad}=0.00020$.}
			\label{fig: H=1a}
		\end{figure}
	\end{center}
\end{widetext}

By transverse period-doubling, the previously ``invisible" waveguide defect mode with in-plane wavevector $\kv_{\parallel}$ distribution peaked at points $\kv_{\parallel}=(\pm \pi/a, \pm \pi/a)$ becomes accessible to the incident light with $k_y=0$, as shown in Fig. \ref{fig: H=1,modes and binding} (a). The real space E-field profile, shown by $\mathrm{E}$-field in the $xy$ plane at the pillar center and  in the $xz$ plane at $y=-0.5a$ in Fig. \ref{fig: H=1,modes and binding} (a), exhibits roughly opposite amplitudes between the A- and B-sublattices for the fields located at the two thin nanopillars of different sizes within each supercell of the line defect. This $180^{\circ}$ phase shift within the defect mode causes destructive wave interference for radiation, and a high $Q$-factor of $11603.9$. Due to the narrow width of the photonic crystal, light leaks into the sidewalls of the flow channel. By increasing the number of unit cells in the PC, the $Q$-factor can be increased to $72571$ (Fig. \ref{fig: H=1a} (c)). We define the radiation fraction $f_\mathrm{rad}$ (listed on the in-plane $\mathrm{E}_{\kv_{\parallel}}$ plot) as the ratio of energy available for radiation $\mathrm{I}_\mathrm{rad}$ inside the lightcone to the total energy $\mathrm{I}_\mathrm{tot}$ stored in the defect mode:
\begin{equation}
f_\mathrm{rad}\equiv \frac{\mathrm{I}_\mathrm{rad}}{\mathrm{I}_\mathrm{tot}}=\frac{\sum_{|\mathbf{k}_{\parallel}| < n_\mathrm{fluid}\cdot \omega/c}
	|\mathrm{E}_{\mathbf{k}_{\parallel}}|^2}{\sum_{\mathbf{k}_{\parallel}}
	|\mathrm{E}_{\mathbf{k}_{\parallel}}|^2}.
\end{equation}
 Components satisfying $|\mathbf{k}_{\parallel}|<
n_{\mathrm{fluid}}\cdot \omega/c$  fall inside the light cone and can couple to propagating radiation modes in the surrounding medium\cite{Q-factor}, while components outside the light cone remain vertically confined. Since the $k_x$ resolution in the in-plane $\kv_{\parallel}$-space is inversely proportional to the chip width in the light propagation $x$-direction, we choose a larger chip width $\mathrm{L}_x=22.86a$, with with eleven silicon blocks from the defect region to the chip end, to improve the accuracy. The resolution $\Delta k_x= 2\pi/\mathrm{L}_x=0.275\; [\pi/a]$ is made much smaller than the lightcone radius $k_\mathrm{light}=n_\mathrm{fluid} \cdot \omega/c= 2.62\; [\pi/a]$. The estimated radiation fraction is $0.0002$, much smaller than $0.0153$ for the $k_y=0$ waveguide mode \cite{ZhaiPB26}, accessible without period doubling.

By varying the in-plane length differences $\left(w_s \times (d\pm dh)\right)$ between the thin silicon nanopillars of row A and row B in the defect region, the magnitude of coupling of incident light to the $k_y=\pi/a$ waveguide mode and the amount of radiative loss can be controlled. For small $dh$, the coupling strength is proportional to the asymmetry parameter $\delta \equiv dh/h$. As this structure asymmetry is reduced, the coupling of the incident plane wave to the very weakly radiating mode is reduced and the transmission decreases.

With a larger structure asymmetry, transmission is enhanced but the portion of low-$\kv_\parallel$ components of the defect mode within the light cone is increased. This reduces the radiative amplitude cancellation $|\mathrm{E}(\rv_\mathrm{A})+\mathrm{E}(\rv_\mathrm{B})|$ and the extent of desctructive wave interference, leading to lower $Q$-factor. 

There are conceptually similar mechanisms involving in-plane structural asymmetry in metasurfaces\cite{Asymmetry1,Asymmetry2} and PC slabs\cite{Asymmetry-PC,Asymmetry-PB,Bandfolding15,Bandfolding23,Bandfolding21,Bandfolding26} exhibiting sharp high-$Q$ resonances, commonly referred to as quasi-bound states in the continuum (``quasi-BICs"). The radiative decay rates and thus the optical lifetimes of these quasi-BICs are determined by the strength of the symmetry-breaking perturbation, leading to $Q$-factors that increases rapidly as the structure approaches the symmetric limit \cite{Asymmetry-PB, Selectionrule}.

In Chip I, the asymmetry $dh=0.1a$ is small, leading to a high $Q$-factor of the order $10^4$ despite a very short chip width $\mathrm{L}_x=4.86a$. As the chip width $\mathrm{L}_x$ in the light propagation $x$-direction is enlarged, i.e. widening the entire flow channel, the $Q$-factor of the defect mode increases and saturates to its intrinsic radiation limit $Q_\perp \sim 72571$. With a larger asymmetry $dh=0.2a$, a higher transmission ($\sim 0.36$) is achieved with the same chip width but a reduced $Q$-factor ($4856.9$) due to a stronger vertical radiation loss. Its larger radiation fraction ($0.00061$) leads to a smaller intrinsic radiation limit ($\sim 15394$), as shown in Fig. \ref{fig: H=1a} (c).

\begin{widetext}
	\begin{center}
		\begin{figure}[htbp]
			\centering
			\vspace{-0.2in} 
			\begin{minipage}[b]{0.45\textwidth}
				\subfigure[]{\hspace{0.2in}\raisebox{0.3in}{\includegraphics[width=\textwidth]{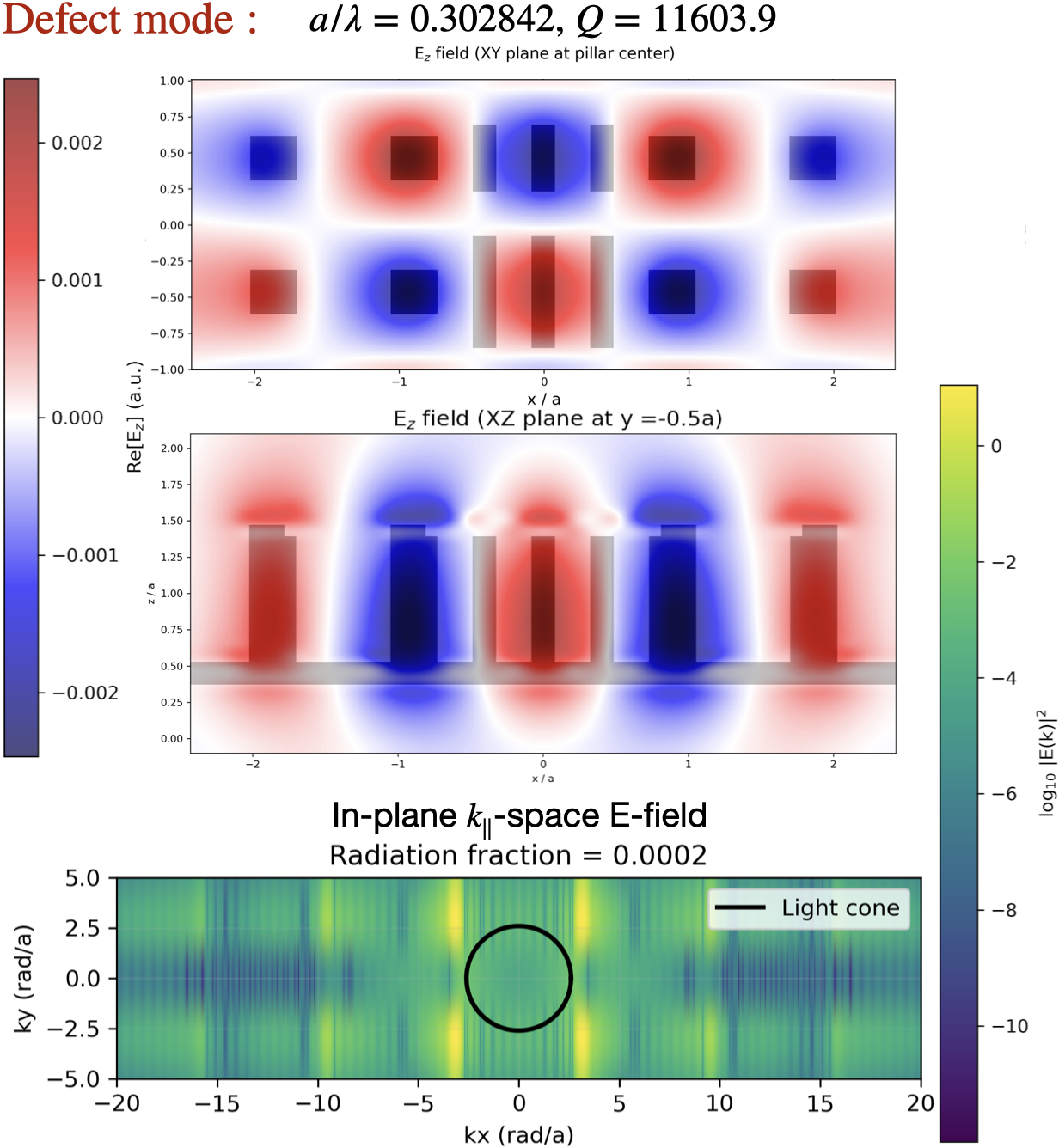} }}
			\end{minipage}
			\hspace{0.01\textwidth}
			\begin{minipage}[b]{0.5\textwidth}
				\centering
				\subfigure[]{\raisebox{-0.1in}{ \includegraphics[width=0.75\textwidth]{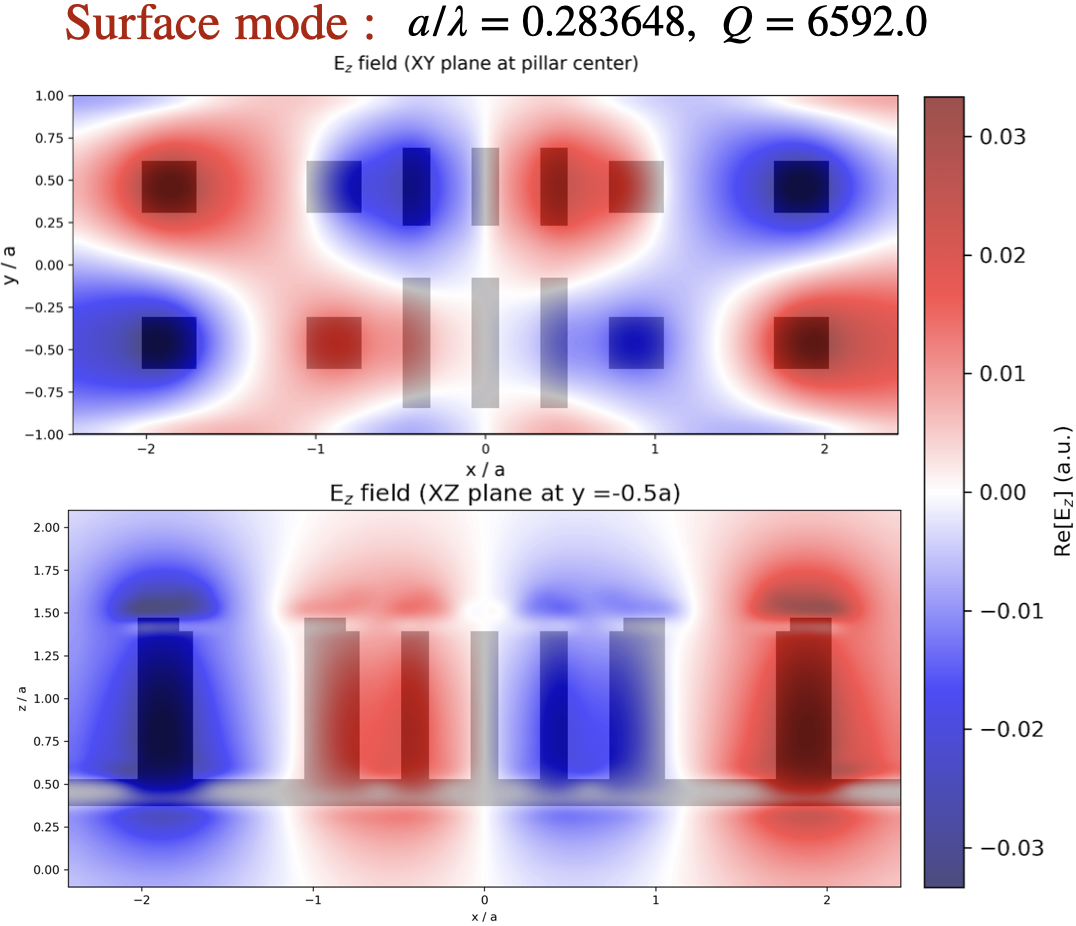} }} \\ 
				\subfigure[]{\raisebox{-0.1in}{ \includegraphics[width=0.95\textwidth]{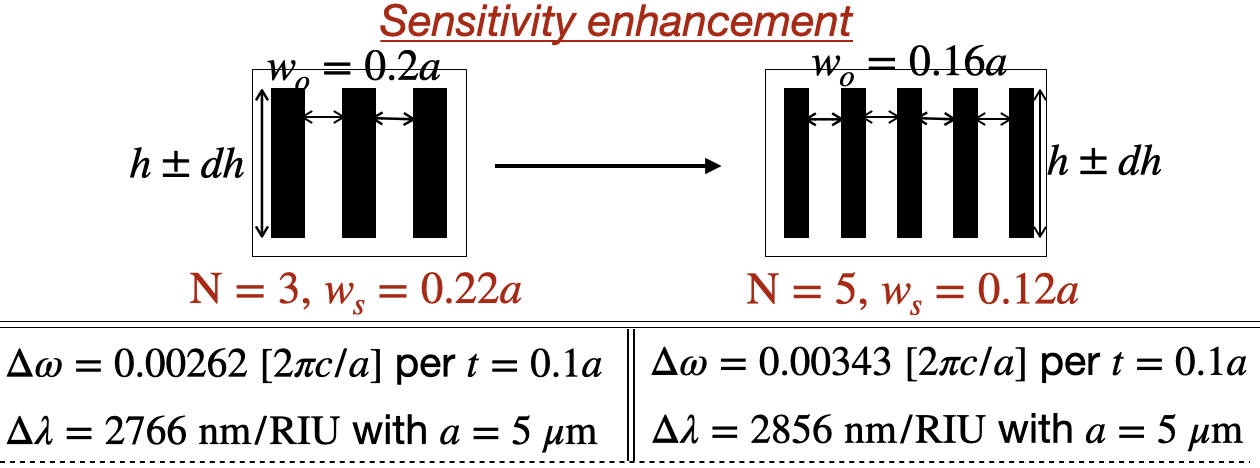} }}
			\end{minipage}
			
			\vspace{0in} 
			
			\hspace{-0.15in}
			\subfigure[]{
				\hspace{-0.2in}\raisebox{-0.2in}{\includegraphics[width=0.5\textwidth]{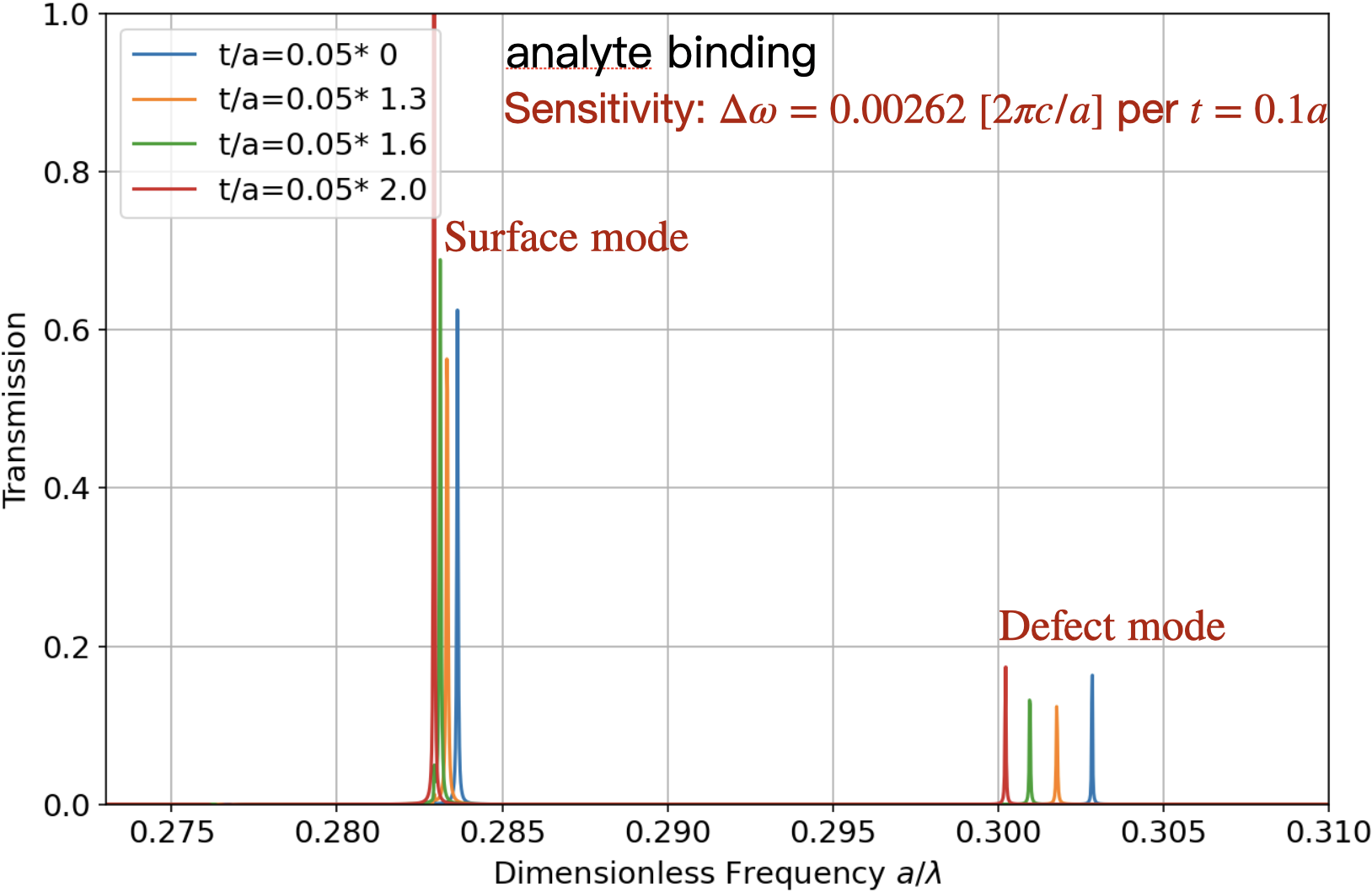}}}
			\hfill	
			\subfigure[]{
				\hspace{0in}\raisebox{-0.2in}{\includegraphics[width=0.5\textwidth]{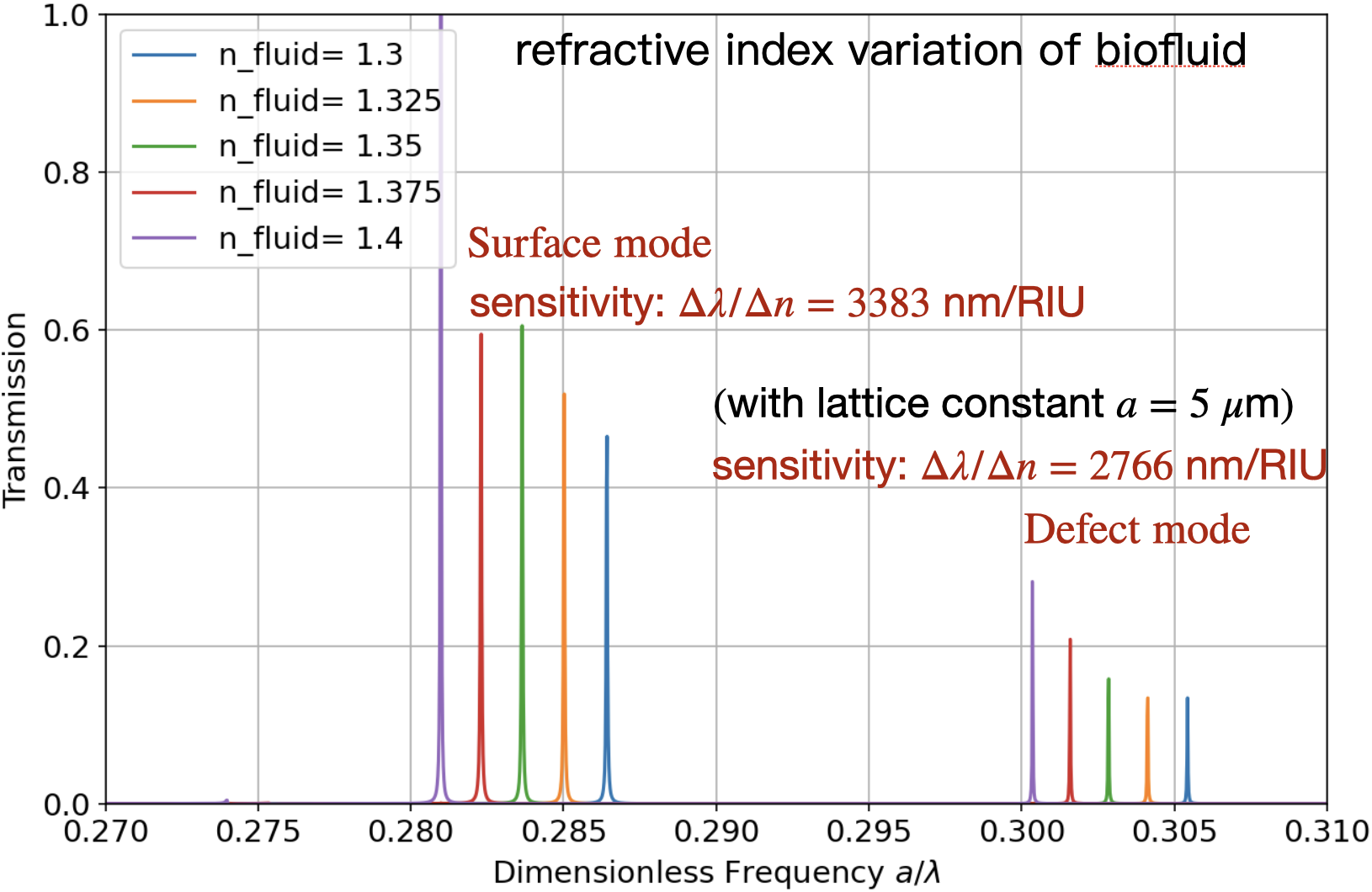}}}
			\caption{(a) E-field profile of the defect mode in the $xy$ plane at the pillar center and in the $xz$ plane at $y=-0.5a$. The in-plane wavevector $\kv_{\parallel}$-space distribution has peaks at $\kv_{\parallel}=(\pm \pi/a, \pm \pi/a)$ with an extremely small tail of low-$|{\bf k}_{\parallel}|$ components lying within the light cone. The estimated radiation fraction is $f_\mathrm{rad}=0.00020$. (b) E-field profile of the left-right anti-symmetric surface mode in the $xy$ plane at the pillar center and in the $xz$ plane at $y=-0.5a$, is spatially peaked near the boundaries of the chip. (c) Frequency shift of transmission peaks with respect to thickness increment of analyte bound in the defect region. The surface mode shifts little with analyte binding. The defect mode is very sensitive to analyte binding, and the sensitivity measured in terms of frequency shift is about $0.00262$ $[2\pi c/a]$ per analyte-thickness $t=0.1a$. (d) Wavelength shift with respect to refractive index variation of the background biofluid. Both waveguide and surface modes are sensitive to this change. The sensitivity for the defect mode in terms of transmission wavelength shift is about $2766 \mathrm{nm/RIU}$ and about $3383$ nm/RIU for the surface mode, with a choice of lattice constant $a = 5\;\mu\mathrm{m}$.}
			\label{fig: H=1,modes and binding}
		\end{figure}
	\end{center}
\end{widetext}

By functionalizing all the surfaces of the thin silicon nanopillars within the defect region  and two neighbouring silicon pillars for analyte-binding (with analyte refractive index $n=1.45 \;(\varepsilon=2.1025)$),
this architecture simultaneously achieves a high sensitivity and low limit-of-detection. 

Transverse period-doubling of the waveguide defect also allows coupling of incident light to an efficacious high-$Q$ surface mode within the PC. Its resonant frequency and $Q$-factor are shown in the transmission spectrum in Fig. \ref{fig: H=1a} (b). Its electric field profiles in the $xy$ plane at the pillar center and in the $xz$ plane at $y=-0.5a$  are shown in Fig. \ref{fig: H=1,modes and binding} (b). This is valuable for sensing the background biofluid composition carrying unattached biomolecules.

Fig. \ref{fig: H=1,modes and binding} (c) and (d) show the transmission response to the analyte-binding and to background biofluid refractive-index variation. For analyte binding in the waveguide defect region, the sensitivity of the waveguide transmission peak is $\Delta \omega=0.00262\;[2\pi c/a]$ per analyte-thickness $t=0.1a$. Since the surface mode is peaked near the walls of the flow channel and only couples evanescently to the analyte binding region, its sensitivity to analyte attachment is considerably less. On the other hand, both modes are sensitive to the refractive index (RI) variation of the background biofluid. The sensitivity of the central waveguide defect mode is $\Delta \lambda =2766$ nm per refractive index unit (RIU)), whereas the surface mode exhibits a higher sensitivity of $\Delta \lambda =3383$ nm/RIU, for $a=5\;\mu$m. Both of these sensitivities are linearly proportional to the unit-cell size $a$. The full-width at half-maximum (FWHM) of the waveguide defect mode is $\Delta \omega^{\text{FWHM}}=\omega/Q=2.61 \times 10^{-5}\; [2\pi c/a]$, which for $a=5\;\mu$m becomes $\Delta \lambda^{\text{FWHM}}=1.423$ nm. The limit-of-detection for the defect mode analyte coating is $\delta t^{\text{lim}}/a=9.96 \times 10^{-4}$. The background fluid limit-of-detection is $\delta n^{\text{lim}}=5.14 \times 10^{-4} $RIU for the defect mode and $\delta n^{\text{lim}}=7.90 \times 10^{-4}$ RIU for the surface mode.

By increasing the number of strips in each row to $\mathrm{N}=5$, reducing the strip width to $w_s=0.12a$, and reducing the spacing between two nearest thin nanopillars to $w_o=0.16a$, the sensitivity of the waveguide mode is improved to $\Delta \omega=0.00343\;[2\pi c/a]$ per analyte-thickness $t=0.1a$. Its other biosensing features are listed in the second row of the Table \ref{fig: Table}, which reveals slightly better biosensing performance at the cost of reduced transmission.

The pillar height in Chip II is reduced to $0.5a$ to simplify fabrication. In Chip II, shown in Fig. \ref{fig: H=0.5a}, the thickness of the thin silicon backing layer is changed to $b=0.1a$. The $x$-direction strip width is changed to $w_s=0.16a$ to pull the resonance into the middle of the PBG. The transmission spectrum after time-delay filtering with $t_\mathrm{delay}=100\; [a/c]$ is shown in Fig. \ref{fig: H=0.5a} (c). With a shorter pillar height and a thinner backing layer, the previous surface mode discussed in Chip I has a higher frequency outside of the PBG. Electric field profiles of the waveguide defect mode in the $xy$ plane at the pillar center and in the $xz$ plane at $y=-0.5a$, and the overall in-plane $\kv_{\parallel}$ distribution (peaked at $\kv_{\parallel}=(0, \pm \pi/a)$) are given in Fig. \ref{fig: H=0.5a} (b). By functionalizing all the surfaces of the thin silicon nanopillars within the defect region  and the two neighbouring silicon pillars for analyte-binding,
Chip II with $\mathrm{H}=0.5a$ remarkably achieves a higher sensitivity and a lower limit-of-detection than Chip I. Its biosensing features are shown in the third row of the Table \ref{fig: Table}. Chip II has an overall better performance than Chip I, at the cost of lower transmission. By increasing the number of strips in each row to $\mathrm{N}=5$, and reducing the strip width $w_s$ and spacing $w_o$ accordingly, the sensitivity can also be further improved, as shown by the fourth row in the Table \ref{fig: Table}.
\begin{widetext}
	\begin{center}
		\begin{figure}[htbp]
			\vspace{-0.1in}
			\subfigure[]{   \hspace{-0.2in}\raisebox{0in}{\includegraphics[width=0.5\textwidth]{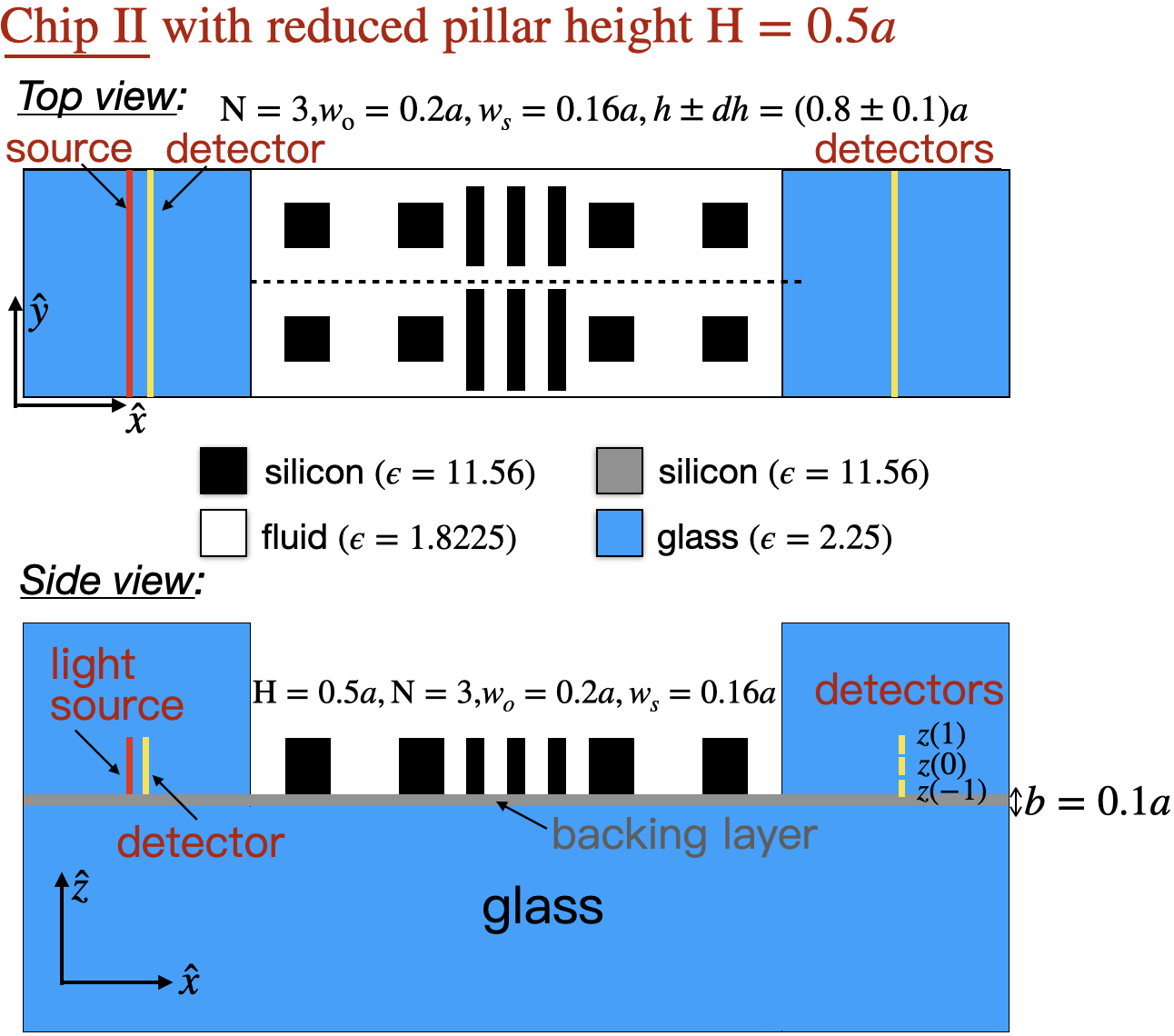}}}
			\subfigure[]{   \hspace{-0.05in}\raisebox{-0.2in}{\includegraphics[width=0.44\textwidth]{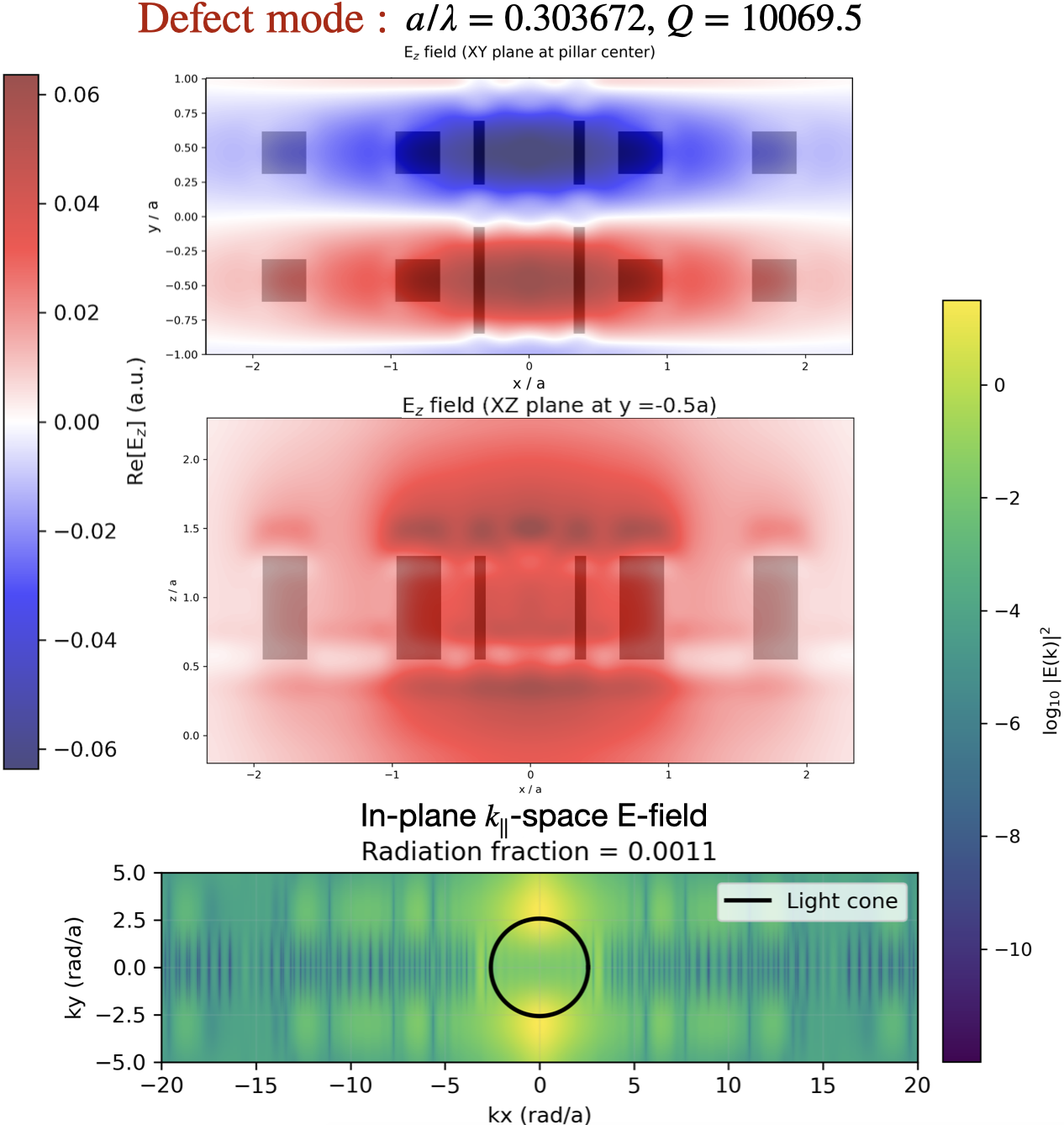}}}			
			\hspace{0in}
			\subfigure[]{
				\hspace{-0.3in}\raisebox{-0.1in}{\includegraphics[width=0.48\textwidth]{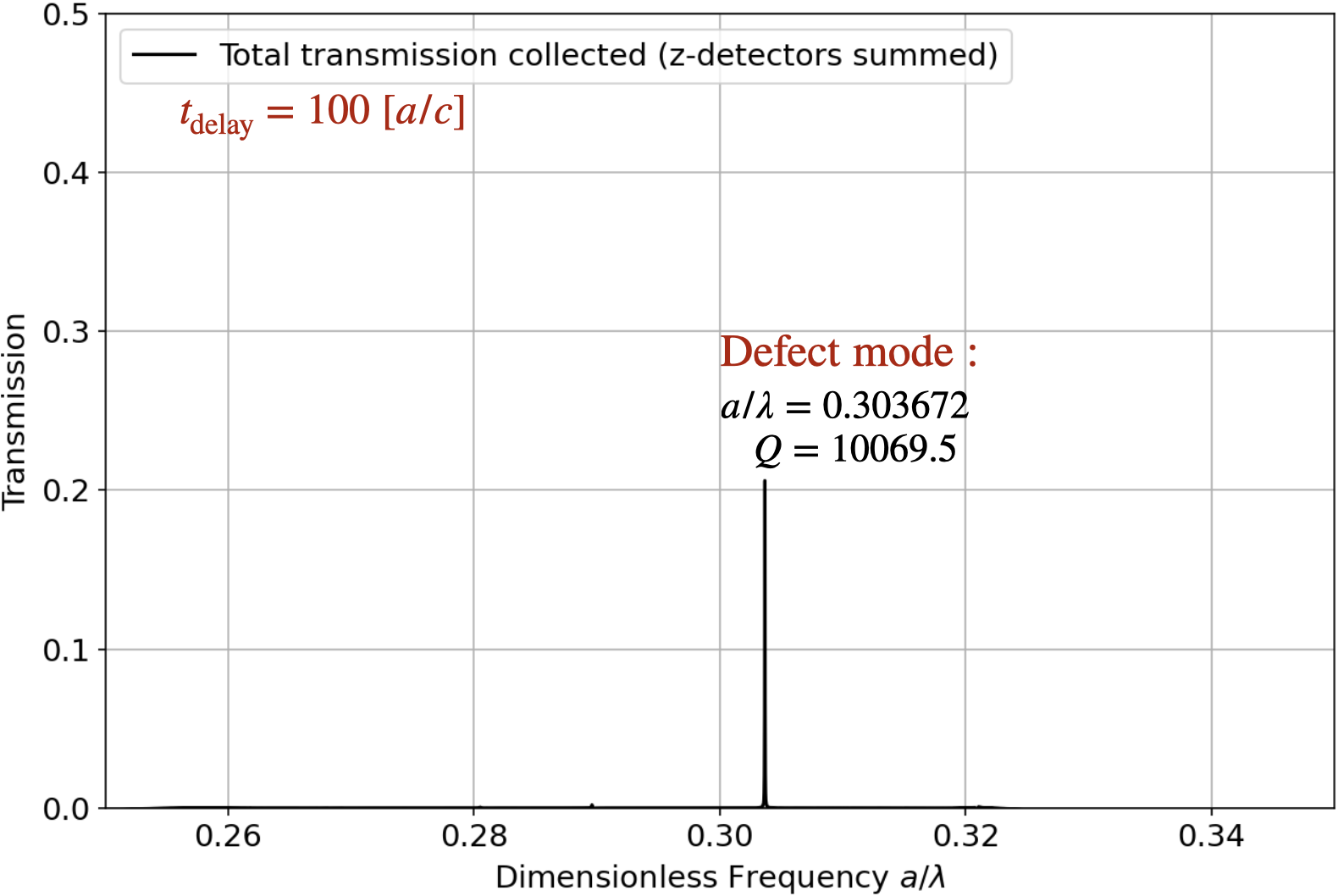}}}
			\hspace{0.15in}
			\subfigure[]{\raisebox{-0.1in}
				{\includegraphics[width=0.5\textwidth]{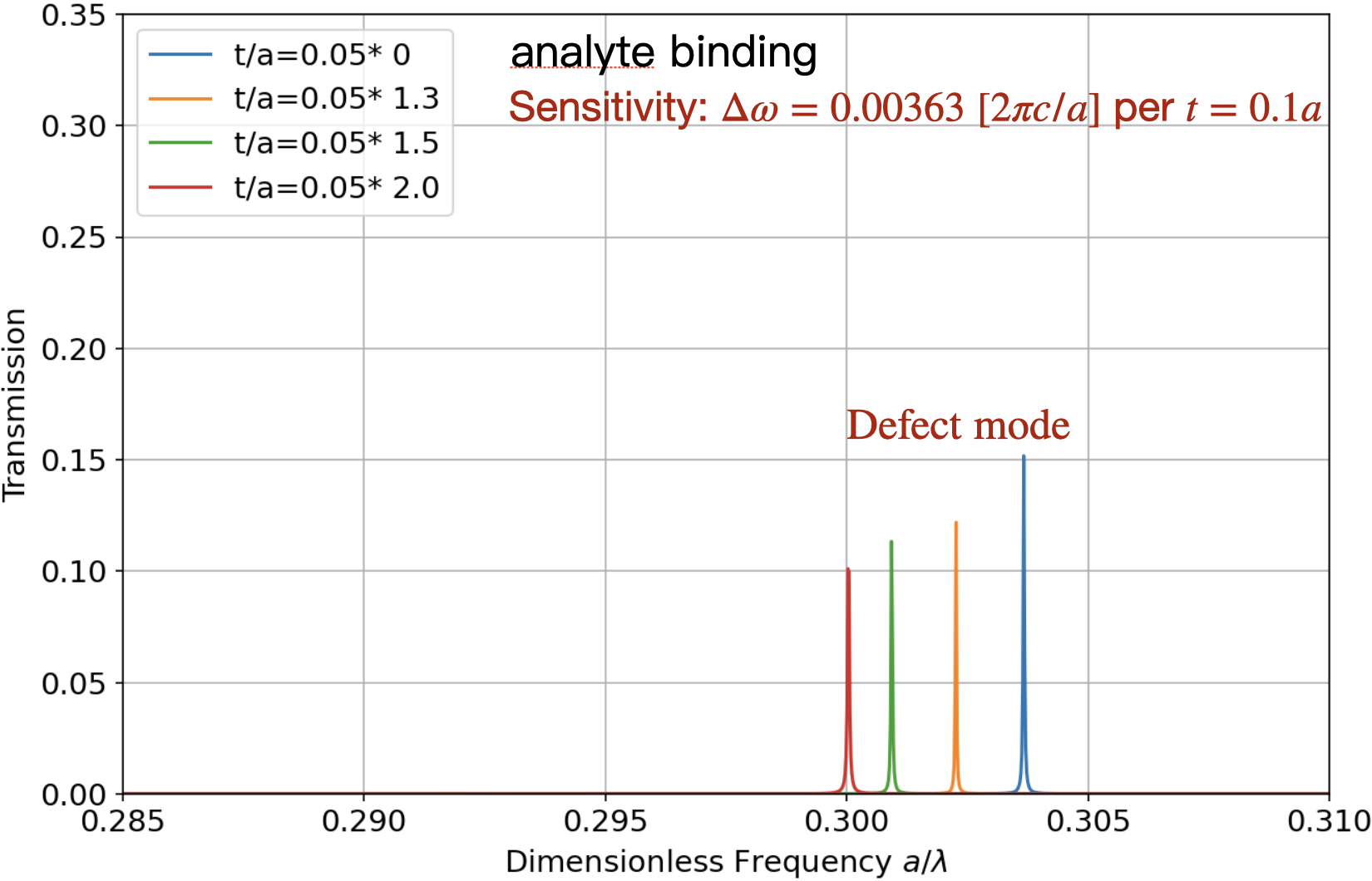}}}
			\caption{Chip II with reduced pillar height $\mathrm{H}=0.5a$. (a) A period-doubled line defect in the middle of the PC slab placed upon a thin silicon backing layer of thickness $b=0.1a$. The in-plane cross-section of each thin silicon pillar in the defect region is $\left(w_s \times (h\pm dh)\right)=\left(0.16a \times (0.8\pm 0.1)a \right)$. The spacing between two nearest thin nanopillars within the defect region is $w_o=0.2a$. (b) E-field profile of the defect mode in the $xy$ plane at the pillar center and in the $xz$ plane at $y=-0.5a$, and Fourier transformed into the in-plane wavevector $\kv_{\parallel}$-space. The defect waveguide mode peaks at $\kv_{\parallel}=(0, \pm \pi/a)$ with a small tail of low-$|{\bf k}_{\parallel}|$ components lying within the light cone. The estimated radiation fraction estimated is $f_\mathrm{rad}=0.0011$. (c) Transmission spectrum after time-delay filtering with a delay time $t_\mathrm{delay}=100\;[a/c]$. With a small asymmetry $dh=0.1a$, the defect mode retains a very high quality factor $Q=10069.5$ that can be greatly increased by widening the PC chip in the $x$-direction. (d) Frequency shift of the transmission peak with respect to thickness increment of analyte bound in the defect region. The sensitivity measured in terms of frequency shift is about $0.00363\;[2\pi c/a]$ per analyte-thickness $t=0.1a$.}
			\label{fig: H=0.5a}
		\end{figure}
	\end{center}
\end{widetext}

Shorter pillar heights ($\mathrm{H}<0.5a$) lead to larger $z$-component momenta $k_z \sim \pi/\mathrm{H}$ within the waveguide defect mode. This pushes the resonant frequency of the defect mode away from the gap center for reasonable values of parameters $w_o$ and $w_s$. A third chip with pillar height $\mathrm{H}=0.25a$ is given, with its parameters and biosensing performance listed in Table \ref{fig: Table}. The asymmetry is increased to $dh=0.2a$ (since $dh=0.1a$ leads to an unacceptably low transmission ($<0.01$)). This has a $Q$-factor of $2924.4$ and a slightly higher resonant frequency due to larger $k_z$ components required for vertical light confinement. Increasing the thickness $b$ of the silicon backing layer pulls down the resonant frequency at the cost of an even lower $Q$-factor.  This third chip has a worse performance than chips I and II due to stronger radiation loss and exhibits very low transmission.
\begin{widetext}
	\begin{center}
   \makeatletter\def\@captype{table}\makeatletter
	\includegraphics[width=0.85\textwidth]{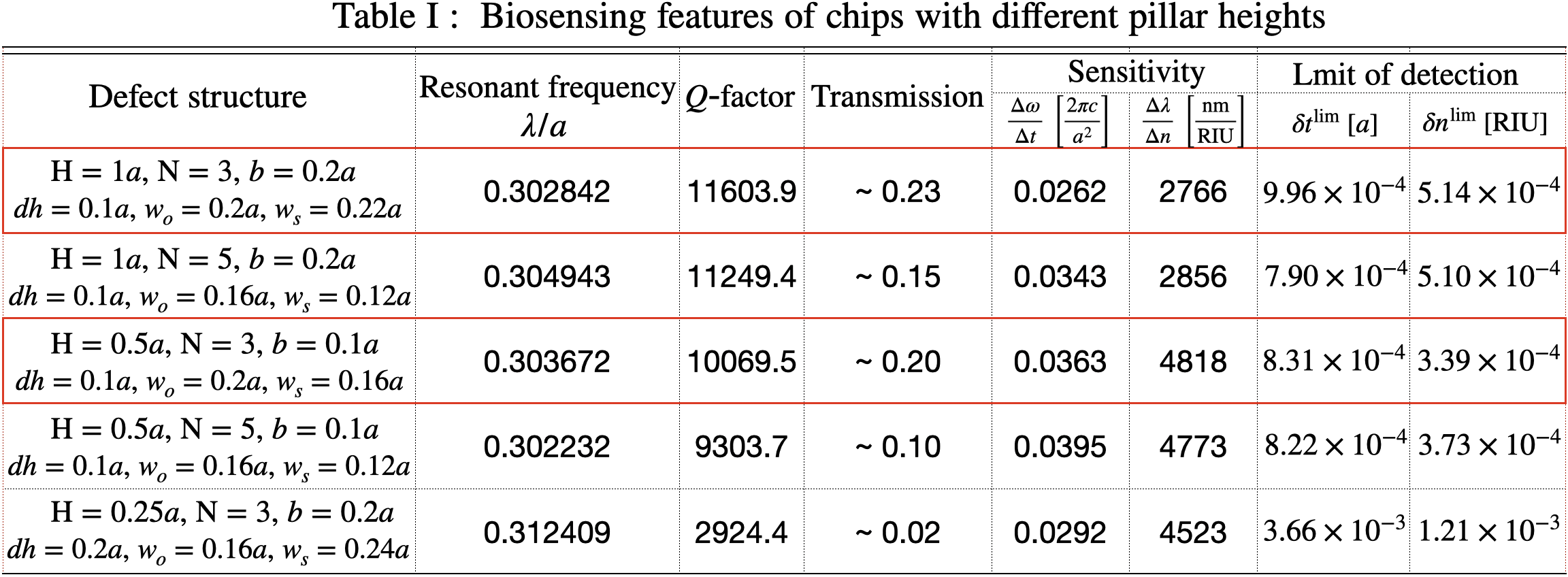}
	\caption{Biosensing features of chips with different pillar heights. The first and third rows framed in red are Chip I and Chip II shown in Fig. \ref{fig: H=1a} and \ref {fig: H=0.5a} respectively. Chips with an increased number $\mathrm{N}=5$ of thin nanopillars in each row within the defect region, both for pillar height $\mathrm{H}=1a$ and $\mathrm{H}=0.5a$, have a higher sensitivity, but a reduced transmission compared with that with $\mathrm{N}=3$. With reducing pillar heights, the transmission decreases due to a stronger radiation loss. For pillar height $\mathrm{H}=0.25a$, the transmission is unacceptably low even with an increased asymmetry $dh=0.2a$. We have chosen the parameters in each case to make the resonant frequency of the defect mode near the center of the 2D PBG. However, the smallest pillar height $\mathrm{H}=0.25a$ causes the resonant frequency to be pushed away from the gap center, for reasonable choices of parameters. Chips with pillar height $\mathrm{H}=0.5a$ show the best biosensing performance with both the highest sensitivity and the lowest limit of detection, with moderate transmissions.}
\label{fig: Table}		
\end{center}
\end{widetext}

In summary, we have presented high-performance PBG-based optical biosensors in 3D with light trapped by period-doubled waveguide defects within a 2D PBG. By
replacing the simple square dielectric pillars with fragmented
dielectric thin nanopillars in the defect region, analyte binds in regions where
the optical field strength is high, enabling high sensitivity to analyte binding. With period doubling along the defect waveguide (and transverse to the light propagation), a novel waveguide mode with features reminiscent of "quasi-bound states in the continuum" is made accessible to incident light. This mode has remarkably weak radiative leakage into the vertical direction due to  destructive wave interference effects. This enables $Q$-factors in the range of $10^4$-$10^5$ depending on the number of photonic crystal unit cells within the biofluid flow channel.

Our photonic crystal biosensor design simultaneously exhibits very high sensitivities and very low limits of detection, the most difficult trade-off to overcome in traditional designs. By using long defect-waveguide analyte attachment regions along the length of the biofluid flow, more rapid analyte binding is realized compared to point defect designs. Lower photonic crystal pillar heights facilitate fabrication at the cost of reduced transmission signal.

Our biosensor design is fully scalable with respect to the choice of lattice constant. We illustrated our design paradigm using $a=5\;\mu$m, making fabrication easier than for a small lattice constant such as $a=0.5\;\mu$m. The larger structure also simplifies the process of functionalizing specific silicon surfaces for analyte attachment. The availability of suitable light sources may also influence the actual choice of lattice constant. For example, reducing the unit cell size from $5$ to $2.5$ microns would reduce the light source wavelengths proportionately. Future work may involve extending the current designs to achieve simultaneous detection of multiple disease-marker analytes from a single biofluid sample. This can be achieved by extending the width of the photonic crystal to include more than one defect waveguide and functionalizing separate waveguides with different aptamers to bind different proteins. Our demonstration of very high sensitivities and very low limits of detection, using a short silicon pillar design, provides a springboard for such future studies.

\acknowledgments
This work was supported by the Natural Sciences and Engineering Research Council of Canada.

\end{document}